\documentclass[12pt]{article}
\usepackage{amsfonts} 
\usepackage{a4}
\usepackage{epsfig}
\usepackage{latexsym}
\begin{document}

\title {Vacuum energy of a massive scalar field in the presence of a semi-transparent cylinder}
\author{Marco Scandurra \thanks{e-mail: scandurr@itp.uni-leipzig.de} \\
  Universit\"at Leipzig, Fakult\"at f\"ur Physik und Geowissenschaften \\
  Institut f\"ur Theoretische Physik\\
  Augustusplatz 10/11, 04109 Leipzig, Germany} \maketitle

\begin{abstract}

We compute the ground state energy of a massive scalar field in the background of a cylindrical shell whose potential is given by a delta function. The zero point energy is expressed in terms of the Jost function of the related scattering problem, the renormalization is performed with the help of the heat kernel expansion. The energy is found to be negative for attractive and for repulsive backgrounds as well.

\end{abstract}

\section{Introduction}

The response of vacuum to classical external fields or constraints has been studied since more than fifty years. The manifestation of this response are macroscopic forces exerted on boundaries or surfaces where the potential is concentrated. The most famous example is the Casimir effect \cite{Casimir1948}, which has been recently measured in laboratory with a precision of nearly $1\%$ \cite{Mohideen}.  Notwithstanding the large size of theoretical work and the experimental efforts  many question remain still open. We have in fact by this far no exact knowledge of the law governing the sign of the Casimir energy, how exactly the features of the boundary and its geometry influence the response of vacuum. An investigation on the cylindrical geometry has begun in \cite{Milton-cyl} after the suggestion that a cylinder, as a kind of intermediate shape between the parallel plates and the  sphere, could possess a zero Casimir stress \cite{pazzo}. However paper \cite{Milton-cyl} showed that a perfectly conducting cylinder in the electromagnetic vacuum has a negative Casimir energy. A study of the dielectric cylinder has been recently carried out in \cite{Nesterenko} and in \cite{Klich}, with the interesting result of a vanishing vacuum energy in the dilute case. 
  Some light could be further shed into the problem with the study of other quantum  fields and other types of background potentials. In this paper we investigate  a cylindrical shell with radius $R$, having a delta function $\delta(r-R)$ as a potential profile. A delta function  is an idealization of a strongly concentrated potential. Unlike the case of Dirichlet boundary conditions the field is continuous on the surface of the shell, therefore the delta function represents a model for a boundary which is not completely hard, but which  becomes transparent at high frequencies. This model can be considered as the ``scalar''  version of the dielectric background. A non singular potential would be, of course, more realistic under the physical point of view, but the calculations would be considerably more complicated. A calculation with the delta function   has been already carried out for a spherical shell \cite{Marco1} and for a magnetic flux tube in the vacuum of spinor  and scalar fields \cite{Marco2}. We hope that our work can contribute in the understanding of vacuum as a fundamental  aspect of quantum field theory. The cylindrical shell could, besides, find physical applications in the calculation of the quantum corrections to vortices in QCD or in the electroweak theory. Another interesting perspective is in the recent discovery of the so called nanotubes \cite{nanotubes1}, \cite{nanotubes2}, which are large carbon molecules generated in laboratory offering the intriguing possibility of measuring quantum effects on small cylindrical objects. It is also of interest the comparison of the problem  of the cylindrical shell with the analogous problem of the magnetic flux tube \cite{Marco2}. 
The organization of this paper is as follow: in the next section we collect the basic formulas necessary for the calculation and we present the renormalization procedure, in section 3 and 4 the major part of the analytical calculation is performed. In section 5 the numerical results are given and in section 6 these results are briefly discussed.

\section{Vacuum energy in terms of the Jost function and renormalization }

Let us work  with a real massive scalar field $\phi (\vec{x},t)$ with mass $m$ and let us quantize it in the background of a cylindrical potential. The field equation in cylindrical coordinates $r,\varphi, z$, after separation of the variables, reads
\begin{equation}
\left(p_0^2-m^2-p_z^2-\frac {l^2}{r^2}-V(r)+
\frac 1r\partial_r +\partial_r^2\right)\phi_l(p_0,p_z,r)\ \ =\ 0 \ ,
\end{equation}
where $p_\mu $ is the momentum four vector being $p_z$ its component along the longitudinal axis of the cylinder and $l$ is the angular momentum quantum number. $V(r)$ is the background potential given by 
\begin{equation}
V(r)\ =\ \frac{\alpha}{R}\delta(r-R)\ ,
\end{equation}    
it represents an infinitely thin cylindrical shell whose profile is a delta function. The shell has a circular section of radius $R$ and it extends from $-\infty$ to $+\infty$ in the $z$ direction. $\alpha$ is a dimensionless parameter giving the strength of the  potential.

Considering eq.(1) as a scattering problem, we choose the ``regular solution''  which is given by
\begin{equation}
\phi(r)\ =\ J_l (kr)\Theta(R-r)+ \frac 12 (f_l(k)H^{(2)}_{l}(kr)+f^*_l(k)H^{(1)}_{l}(kr))\Theta(r-R) \ ,
\end{equation}
where $k=\sqrt{p_0^2-m^2-p_z^2}$, $J_m (kr)$ is a Bessel function of the first kind, $H^{(1)}_l(kr)$ and $H^{(2)}_l(kr)$ are the Hankel functions of the first and of the second kind and the coefficients $f_l(k)$ and $f^*_l(k)$ are a Jost function and its complex conjugate respectively. $\Theta(R-r)$ is a theta function. The field is therefore free in the regions $0<r<R$ and $R<r<\infty$. At $r=R$ the field is continuous and we have the following matching conditions
\begin{equation}
\left\{ \begin{array}{ll}\phi '(R+0) - \phi '(R-0) =\ 
\frac{\alpha}{R}\phi (R)\\ 
\phi(R+0)\ =\  \phi(R-0)\ , \end{array} 
\right.
\end{equation}
where the prime indicates derivative with respect to $r$.\\
The quantum field is in its vacuum state, its energy is given by half of the sum over all possible eigenvalues $\omega_{(n)}$ of the Hamilton operator related with the wave equation (1). We define a regularized vacuum energy
\begin{equation}
E\ =\ \frac {\mu^2 s}{2}\sum_{(n)} \omega_{(n)}^{1-2s}\ ,
\end{equation}
where $s$ is the regularization parameter which we will put to zero after the renormalization, $\mu$  is a mass parameter introduced for dimensional reasons and $(n)$ includes all possible quantum numbers. We will calculate the energy density per unit length of the cylinder given by
\begin{equation}
{\cal E}\ =\ \frac 12 \mu^{2s}\int^{\infty}_{-\infty}\frac{dp_z}{2\pi}
\sum_{(n)} (p_z^2+ \epsilon_{(n)}^2)^{1/2-s}\ ,
\end{equation}
where $\epsilon_{(n)}$ are the eigenvalues of the operator contained in (1) without $p_z$ and with $k=\sqrt{p_0^2-m^2}$. We carry out the integration over $p_z$ and we arrive at
\begin{equation}
{\cal E}\ =\ \frac 14 \mu^{2s}\frac{\Gamma(s-1)}{\sqrt{\pi}\Gamma(s-1/2)}\sum_{(n)}(\epsilon_{(n)}^2)^{1-s}\ .
\end{equation}
Following a well known procedure\footnote{ The procedure is explained with details in \cite{firstball}.}  we transform the sum in (7) into a contour integral and, dropping the Minkowski space contribution, we find
\begin{equation}
{\cal E} \ =\ -\frac 14 C_s\ \sum^\infty_{l=-\infty}\int^\infty_m dk\ (k^2+m^2)^{1-s} \partial_k \ln f_l(ik)\ ,
\end{equation}
 where $C_s=(1+s(-1+2\ln(2\mu)))/(2\pi)$ and $f_l(ik)$ is the Jost function defined in (3) on the imaginary axis. It contains all the information about the background potential under examination. We will find $ f_l(ik) $ explicitly in the following section. The energy defined in (8) is renormalized by direct subtraction of its divergent part
\begin{equation}
{\cal E}_{ren}\ =\ {\cal E}\ -\ {\cal E}_{div}\ 
\end{equation} 
with the normalization condition demanding that the vacuum fluctuations vanish for a field of infinite mass
\begin{equation} 
\lim_{m \rightarrow \infty }{\cal E}_{ren} = 0\ .                       
\end{equation} 
This normalization condition fixes the arbitrariness of the mass parameter $\mu$ (for more details see \cite{Boston}. The condition does not apply to massless fields).
The isolation of ${\cal E}_{div}$ is done with the use the heat-kernel expansion of the ground state energy 
\begin{equation}
{\cal E}\ =\ \sum_j \frac{\mu^{2s}}{32\pi^2}\frac{\Gamma(s+j-2)}{\Gamma(s+1)}m^{4-2(s+j)}A_j\ \ ,\ \ \ \  j=0, \frac 12, 1,...\ ,     
\end{equation}
where the $A_j$ are the heat kernel coefficients related to the background.
Then we define
\begin{eqnarray} 
{\cal E}_{div} &  =  & -\frac{m^4}{64 \pi ^2}\left( \frac 1s + \ln
\frac{4\mu ^2 }{m^2} - \frac 12\right) A_0\ -\frac{m^3}{24\pi^{3/2}}A_{1/2}
\nonumber \\     
                  &     &  +\frac{m^2}{32 \pi ^2}\left( \frac 1s +
\ln\frac{4\mu ^2 }{m^2} - \ 1\right) A_1 \ + \frac{m}{16 \pi^{3/2}}A_{3/2} \nonumber \\ 
                  &     &  -\frac{1}{32 \pi ^2}\left( \frac 1s + \ln \frac{4\mu ^2 }{m^2} - 2\right) A_2\ ,  
\end{eqnarray} 
which includes all the pole terms and all the terms proportional to non-negative powers of the mass.

In order to perform the analytical continuation $s\rightarrow 0$  we split the renormalized vacuum energy (9) into  two parts
\begin{equation}
{\cal E}_{ren}\ =\ {\cal E}_f + {\cal E}_{as},
\end{equation}
with 
\begin{equation}
{\cal E} _f\ =\  - \frac 14 C_s \sum_{l=-\infty}^\infty\int^\infty_m dk [k^2 -m^2]^{1-s} \frac{\partial}{\partial k} [\ln f_l(ik)- \ln f^{as}_l(ik)]\end{equation}
and
\begin{equation}
{\cal E} _{as}\ =\  -\frac 14 C_s\sum_{l=-\infty}^\infty \int^\infty_m dk [k^2 -m^2]^{1 -s} 
\frac{\partial}{\partial k}  \ln f^{as}_l(ik)-{\cal E}_{div} .
\end{equation}
Here $f^{as}_l(ik)$ is a portion of the uniform asymptotic expansion of the Jost function which must include as many orders in $l$ as it is necessary to have
\begin{equation}
\ln f_l(ik) -  \ln f^{as}_l(ik)\ = {\cal O} \left( l^{-4}\right)
\end{equation}
in the limit $l\rightarrow \infty$,  $k\rightarrow \infty$ with fixed $l/k$, which is sufficient to let the sum and the integral in (14) converge at $s=0$. The splitting proposed in (13) leaves the quantity ${\cal E} _{ren}$ unchanged, while it permits the substitution $s=0$ in the finite part ${\cal E} _f$.

\section{The Jost function and its asymptotics}

We insert solution (3) into (4) and we find
\begin{equation}
\left\{ 
\begin{array}{ll}
J_{l}(kR)  =  \frac 12 \left[f_l(k) H^{(2)}_l(kR)+f^{\star}_l(k) H^{(1)}_l(kR)\right] \nonumber \\  
\left(\frac 12 \partial_r \left[f_l(k) H^{(2)}_l(kR)+f^{\star}_l(k) H^{(1)}_l(kR)\right]   \right)|_{r=R} =  \frac{\alpha}{R} J_{l}(kR) + \left( \partial_r J_{l}(kr)\right)|_{r=R}\ .
\end{array}
\right. 
\end{equation}
The system can be solved for $f_l(k)$, with the help of the Wronskian determinant of the Hankel functions \cite{Abramowitz}, the result is
\begin{eqnarray}
f_l(k) & = &  -\frac 12 i\pi kR \left( J_{l+1}(kR) H^{(1)}_l(kR)- J_{l}(kR) H^{(1)}_{l+1}(kR)\right.\nonumber \\
       &   &    \left.  -\frac{\alpha}{kR}J_{l}(kR) H^{(1)}_l(kR)\right)\ .
\end{eqnarray}
The corresponding Jost function on the imaginary axis can be written in terms of the modified Bessel I and K functions, again with the help of \cite{Abramowitz}
\begin{equation}
f_l(ik)\ =\ 1+\alpha\ I_l(kR)K_l(kR).
\end{equation}
From the Jost function (19) one arrives at the uniform asymptotic expansions $f^{as+}_l(ik)$ for positive $l$ and  $f^{as-}_l(ik)$ for negative $l$ and  at the asymptotic expansion $f^{as}_0(ik)$ for $l=0$, with the help of  the asymptotics of the Bessel I and K functions for large indices and large arguments available on \cite{Abramowitz}. Since the asymptotic Jost function consists  of three different contributions, the sum over $l$ appearing in (14) and (15) must also be distinguished in three  contributions: a sum over negative $l$, a sum over positive $l$ and a contribution coming from $l=0$. The first two contributions can be summed up analytically in the following way 
\begin{equation}
\sum_{l=-\infty}^{-1}...\ln f^{as-}_l(ik)...\ +\ \sum_{l=1}^{\infty}...\ln f^{as+}_l(ik)...\ =\  \sum_{l=1}^{\infty}...(\ln f^{as+}_l(ik)+ \ln f^{as-}_{-l}(ik))...\ ,
\end{equation}
where the dots represent, for simplicity, the rest of the functions in (14) and (15). Then, eq.(14) and (15) are rewritten in the form
\begin{eqnarray}
{\cal E} _{as} & = &  -\frac 14 C_s\sum_{l=1}^\infty \int^\infty_m dk [k^2 -m^2]^{1 -s} \frac{\partial}{\partial k} \ln f^{as\pm}_l(ik)  \nonumber \\
               &   &  -\frac 14 C_s \int^\infty_m dk [k^2 -m^2]^{1 -s} \frac{\partial}{\partial k} \ln f^{as}_0(ik) -{\cal E}_{div} 
\end{eqnarray}
and
\begin{eqnarray}
{\cal E} _f & = &  -\frac {1}{8\pi} \sum_{l=1}^\infty\int^\infty_m dk [k^2 -m^2] \frac{\partial}{\partial k} \left(2\ln f_l(ik)-\ \ln f^{as\pm}_l(ik)\right) \nonumber\\
            &   &  -\frac{1}{8\pi} \int^\infty_m dk [k^2 -m^2] \frac{\partial}{\partial k} \left(\ln f_0 (ik)-\ln f^{as}_0(ik)\right)\ ,
\end{eqnarray}
where  $\ln f^{as\pm}(ik)=\ln f^{as+}_l(ik)+\ln f^{as-}_{-l}(ik)$ and
we have used the property $f_l(ik)=f_{-l}(ik)$ of eq.(19).

 Taking the logarithm of the uniform asymptotics of the  modified Bessel functions and re-expanding in negative powers of the variable $l$ (see \cite{Marco2} for details on this procedure, see also appendix A) we find
\begin{equation}
\ln f_0^{as}\ =\ \frac{\alpha}{2kR}\ -\ \frac{\alpha^2}{8k^2R^2}\ ,
\end{equation}
\begin{equation}
\ln f^{as\pm}(ik)\ =\ \sum_{n=1}^3\sum_jX_{n,j}\frac{t^j}{l^n}\ ,
\end{equation}
where $t=(1+(kR)/l)^2)^\frac 12 $  and the non-vanishing  coefficients are
\begin{equation}
\begin{array}{l}
X_{1,1}=\alpha\ ,\ X_{2,2}=-\alpha^2/4\ , \\
X_{3,3}=\alpha/8+\alpha^3/12\ ,\ X_{3,5}=-3\alpha/4\ ,\\ X_{3,7}=5\alpha/8\ .
\end{array}
\end{equation}
In this definition we have included 3 orders in $l$ which are sufficient to satisfy condition (16). Substituting (23) and (24) in (21) and (22) we find
\begin{eqnarray}
{\cal E} _{as} & = &  -\frac 14 C_s\sum_{l=1}^\infty \int^\infty_m dk [k^2 -m^2]^{1 -s} \frac{\partial}{\partial k} \left(\sum_{n=1}^3\sum_jX_{n,j}\frac{t^j}{l^n}\right)  \nonumber \\
               &   &  -\frac 14 C_s \int^\infty_m dk [k^2 -m^2]^{1 -s} \frac{\partial}{\partial k}  \left(\frac{\alpha}{2kR}\ -\ \frac{\alpha^2}{8k^2R^2}\right) -{\cal E}_{div} 
\end{eqnarray}
and
\begin{eqnarray}
{\cal E} _f & = &  -\frac {1}{8\pi} \sum_{l=1}^\infty\int^\infty_m dk [k^2 -m^2] \frac{\partial}{\partial k} \left(2\ln f_l(ik)-\sum_{n=1}^3\sum_jX_{n,j}\frac{t^j}{l^n} \right) \nonumber\\
            &   &  -\frac{1}{8\pi} \int^\infty_m dk [k^2 -m^2] \frac{\partial}{\partial k} \left(\ln f_0 (ik)-\frac{\alpha}{2kR}\ -\ \frac{\alpha^2}{8k^2R^2} \right)\ .
\end{eqnarray}
We call the first addend in (27) ${\cal E} _{fl}  $ and the second ${\cal E} _{f0} $, that is ${\cal E} _f={\cal E} _{fl}+{\cal E} _{f0}$. 

\section{Asymptotic part of the energy}

We go forward with an analytical simplification of (26). We call the second addend in (26) ${\cal E}_{as0}$, it can be immediately calculated, giving
\begin{equation}
{\cal E} _{as0}\ =\ -\frac{\alpha m}{8 \pi R}\ -\ \frac{\alpha^2}{64 \pi R^2}\left(\frac 1s +\ln\left(\frac{4\mu^2}{m^2}\right) -2\right)\ .
\end{equation}
A simplification of the first addend in (26) can be achieved with the Abel-Plana formula (appendix B) which transforms the sum over $l$ into an integral. Then, the first addend in (26) turns out to be the sum of three contributions
\begin{eqnarray}
{\cal E}_{as1} & = & \frac{\alpha m^2 }{16\pi}\left(\frac 1s +\ln\left(\frac{4\mu^2}{m^2}\right) -1\right)\ +\ \frac{\alpha^3 }{96\pi R^2}\left(\frac 1s +\ln\left(\frac{4\mu^2}{m^2}\right) -2\right)\nonumber\\ 
               &   &   +\frac{\alpha^2 m}{32 R}\ ,
\end{eqnarray}
\begin{equation}
{\cal E}_{as2}\ =\ \frac{\alpha m }{8\pi R}\ +\ \frac{\alpha^2 }{64\pi R^2}\left(\frac 1s +\ln\left(\frac{4\mu^2}{m^2}\right) -2\right)\ -\ \frac{\alpha}{64\pi m R^3}\ -\ \frac{\alpha^3 }{96\pi m R^3}\ ,
\end{equation}
\begin{eqnarray}
{\cal E}_{as3} & = & -\frac{\alpha}{2\pi R^2}\  h_1(mR)\ -\ \frac{\alpha^2}{32  R^2}\ h_2(mR)+\left( \frac{\alpha}{16 \pi R^2}+\frac{\alpha^3}{24 \pi R^2}\right)\ h_3(mR)\nonumber \\ 
       &  & -\frac{\alpha}{8 \pi R^2}\  h_4(mR)\ +\frac{\alpha}{48 \pi R^2}\  h_5(mR)\ ,      
\end{eqnarray}
where the functions $h_i(x)$ are convergent integrals over $l$ given explicitly in appendix B. In appendix B the reader can also find more details about the derivation of the three contributions (29), (30) and (31). ${\cal E}_{as1}$ and ${\cal E}_{as2}$ contain all the pole terms (all the divergences of the vacuum energy) plus the terms proportional to non-negative powers of the mass (which do not satisfy the normalization condition). All these terms are subtracted and  are used to calculate the heat-kernel coefficients by means of definition (12) for ${\cal E}_{div}$. Below we give the heat kernel coefficients which we calculated up to the coefficient $A_4$ (adding four more orders in $\ln f_l^{as\pm}(ik)$), in the hope that they will be of use for future investigations on the same background 
\begin{equation}
\begin{array}{lclcccl}
A_0 & = & 0                 & , &   A_{1/2} & = & 0 \nonumber \\
A_1 & = &  -2\pi\alpha     & , &   A_{3/2} & = &  \frac{\alpha^2 \pi^{3/2}}{2R} \nonumber \\
A_2 & = &  \frac{\pi \alpha^3}{3 R^2}                & , &   A_{5/2} & = & -\frac{(3\alpha^2 +4\alpha^4)\pi^{3/2}}{192R^3} \nonumber \\
A_3 & = &  \frac{(4\alpha^3+7\alpha^5)\pi}{210 R^4}               & , &  A_{7/2}  & = & -\frac{(81\alpha^2 +120\alpha^4+128\alpha^6)\pi^{3/2}}{24576 R^5}\nonumber \\
A_4 & = &  \frac{(64\alpha^3+52\alpha^5+39\alpha^7)\pi}{16380 R^6}  & . &  &  & 
\end{array}
\end{equation}
One can note how all the integer heat-kernel coefficients depend on odd powers of the coupling constant, while the half-integer coefficients depend on even powers of $\alpha$. The same feature is present in the heat kernel coefficients of a $\delta$-potential spherical shell \cite{Marco1}.

We perform the subtraction of the divergent portion and we obtain the final result
\begin{eqnarray}
{\cal E}_{as} & = & -\frac{\alpha}{2\pi R^2}\  h_1(mR)\ -\ \frac{\alpha^2}{32  R^2}\ h_2(mR)\nonumber \\ 
       &  & +\left( \frac{\alpha}{16 \pi R^2}+\frac{\alpha^3}{24 \pi R^2}\right)\ h_3(mR)\nonumber \\ 
       &  & -\frac{\alpha}{8 \pi R^2}\  h_4(mR)\ +\frac{\alpha}{48 \pi R^2}\  h_5(mR)\ -\ \frac{\alpha}{64 \pi m R^3}\ -\frac{\alpha^3}{96 \pi m R^3}\ .      
\end{eqnarray}

\section{Finite part of the energy and numerical results}

The quantity (27) cannot be analytically simplified. We integrate ${\cal E} _{fl} $ and ${\cal E} _{f0} $ by parts and make the substitution $k\rightarrow k/R$ to get an explicit dependence on $R$
\begin{eqnarray}
{\cal E} _{fl} & = &  \frac {1}{4\pi R^2} \sum_{l=1}^\infty\int^\infty_{mR} dk\ k\  \left(2\ln f_l(ik)|_{k\rightarrow k/R }-\left(\sum_{n=1}^3\sum_jX_{n,j}\frac{t^j}{l^n}  \right)|_{k\rightarrow k/R }\right)\nonumber\\
{\cal E} _{f0} & = &  \frac{1}{4\pi R^2}  \int^\infty_{mR} dk\ k\ \left( \ln f_0 (ik)|_{k\rightarrow k/R }-\frac{\alpha}{2k}\ +\ \frac{\alpha^2}{8k^2} \right)\ .
\end{eqnarray}
For small values of $R$, ${\cal E} _{fl}$ behaves like $R^{-2}$, while  ${\cal E} _{f0} $ has a logarithmic behaviour
\begin{equation}
{\cal E} _{f0} \ \sim\ -\frac{\alpha^2\ln (mR)}{32\pi R^2}\ .
\end{equation}
We note that in the limit $\alpha\rightarrow 0$, the logarithm of the Jost function $f_l(ik)$ can be expanded in powers of $\alpha$, giving (from eq.(19))
\[
\ln (1+\alpha I_l(kR)K_l(kR)) \sim\  \alpha I_l(kR)K_l(kR)\ +\ \frac 12\alpha^2 I_l^2(kR)K_l^2(kR)\ +\ {\cal O}(\alpha^3),
\]
then, the summation over $l$ of the leading term of this expansion could be analytically performed following a method recently proposed in \cite{Klich}. This would  give us a first order approximation of ${\cal E}_{fl}$. However we will restrict us  here  to a  numerical calculation of the sum in (34), which can be performed for small as well as for large values of $\alpha$.
  
To find the asymptotic behaviour of ${\cal E}_{as}$ we rewrite eq.(33) in the following form
\begin{equation}
{\cal E}_{as}\ =\ \frac{1}{2\pi R^2}\left[\alpha w_1(mR)\ +\ \alpha^2 w_2(mR)\ +\ \alpha^3 w_3(m R)\right]\ ,
\end{equation} 
where the functions $w_1(x)$, $w_2(x)$ and $w_3(x)$ are given by
\begin{eqnarray}
w_1(x) & = & \left(-h_1(x)+\frac{1}{8}h_3(x)-\frac 14 h_4(x) +\frac{1}{24}h_5(x)-\frac{1}{32x} \right)\ ,\nonumber\\
w_2(x) & = & \left(-\frac{\pi}{16}h_2(x)\right)\ ,\nonumber\\
w_3(x) & = & \left(\frac{1}{12}h_3(x)-\frac{1}{48x}\right)\ .
\end{eqnarray}
As mentioned above, the functions $h_i(x)$ are quickly converging integrals. The behaviour of ${\cal E} _{as}$ is governed by the functions $w_{1,2,3} (x)$ and by the value of $\alpha$. For $R\rightarrow 0$ we find
\begin{eqnarray}
w_1(x) & \sim & 0.0000868\ +\ {\cal O}(x)\ ,\nonumber\\
w_2(x) & \sim & \frac{1}{16}\ln x\ +\ 0.115\ +\ {\cal O}(x) \ , \nonumber\\
w_3(x) & \sim & \frac{1}{24}\ln x\ +\ 0.00479\ +\ {\cal O}(x) \ .
\end{eqnarray}
Thus, for small values of the radius, ${\cal E} _{as}$ is proportional to $\ln R$. Summing the contribution coming from (35) and (38), we find that the leading term of the renormalized energy, for $R\rightarrow 0$, is 
\begin{equation}
{\cal E} _{ren}\ \sim \ \frac {\alpha ^3\ln (mR)}{48\pi R^2}\ . 
\end{equation}
This result is in agreement with the prediction (section II of paper \cite{master})
\begin{equation}
\lim_{R\rightarrow 0}{\cal E} _{ren} \ \sim\ \frac {A_2}{16\pi^2}\ln (mR)\ .
\end{equation}  
In the limit $R\rightarrow \infty $ the behaviour of the renormalized energy is determined by the first non-vanishing heat kernel coefficient after $A_2$. From (32) and eq.(11) we arrive at
\begin{equation}
\lim_{R\rightarrow \infty}{\cal E} _{ren} \ \sim\ -\frac {\alpha^2}{2048 mR^3}\ -\frac {\alpha^4}{1536 mR^3}\ +{\cal O}\left(\frac{1}{R^4}\right)\ .
\end{equation} 

We have numerically evaluated the quantities ${\cal E} _{as}$, ${\cal E} _{fl}$, ${\cal E} _{f0}$ and ${\cal E} _{ren}$ as functions of $R$, fixing the values of the mass to $1$. It turned out to be necessary to sum  20 terms in the variable $l$ and to integrate up to 1000 in the variable $k$  to obtain ``stable'' numerical values for the energy.
Below we give the plots of the various contributions to the vacuum energy and
the complete renormalized  energy for different values of the potential
strength. For $\alpha<0$ we found the renormalized vacuum energy to posses a
small imaginary part becoming larger when $\alpha$ approaches $-\infty$.
Particle creation accounts for this contribution. It starts when the attractive potential of the shell becomes over-critical, that is when ${\cal \epsilon}<-m<0$, where ${\cal \epsilon}$ is the energy of the bound state; in this case the effective action of the system acquires an imaginary part\footnote{For the theoretical foundations of this phenomenon see \cite{Schwinger}. In the context of vacuum energy the appearance of an imaginary part in the renormalized energy was  observed for instance in \cite{BordagKirstenHellmund}.}, however a detailed discussion of this aspect of the theory is beyond the purpose of the present paper. It should only be said that in the plots traced for negative values of $\alpha$ (Fig. 4 and 5) the energy is to be intended as real part of.

\section{Conclusions}

We calculated the vacuum energy of a scalar field in the background of a
cylindrical semi-transparent shell. The formulas for the energy density per
unit length of the shell are given by equations (33) and (34). The heat kernel
coefficients (32) for a cylindrical potential containing a delta function, are also a relevant part of the results. A discussion of the sign of the
vacuum energy is possible by means of the data given in the numerical
section of this paper. The energy is found to be negative in the background of repulsive potentials ($\alpha>0$) for every finite value of the radius (Fig.  3).  This conclusion sets a closeness between the model studied
here and that of a conducting cylinder \cite{Milton-cyl} and of a dielectric
cylinder \cite{Nesterenko}  in the electromagnetic vacuum. The latter models possess also
a  negative Casimir energy. There is also a resemblance with the $\delta$-potential spherical shell investigated in paper \cite{Marco1}, where a negative energy was observed for large repulsive potentials. 

 In the background of attractive potentials ($\alpha <0$) the energy is negative on almost all the $R$-axis. As we see in Fig. 5, the energy becomes positive for very small values of the radius. This happens when the logarithmic term in (39) compensates for the term proportional to $-\alpha^2$ in (41) which is dominant for small values of $\alpha$ in the region $R>1$. The thin region of positive asymptotically increasing energy appearing in Fig. 5 is reasonably out of physical applicability. 
For both repulsive and attractive potentials the renormalized vacuum energy of the semi-transparent cylinder goes logarithmically to $\pm \infty$ in the limit $R\rightarrow 0$. This logarithmic behaviour was also found in the
semi-transparent spherical shell \cite{Marco1},  however it was not found in the $\delta$-potential flux tube \cite{Marco2}, which has
many feature in common with the background investigated here. This can be explained by means of the heat-kernel coefficient $A_2$, which is a non zero coefficient here as well as in \cite{Marco1}. It is also interesting to note how in expression (32) for the coefficient $A_2$ the contributions proportional to $\alpha$ and $\alpha^2$ have cancelled and only a term proportional to $\alpha^3$ is present. The cancellation of the lower powers of the potential strength was also observed in \cite{Marco1} and in paper \cite{dielectric}, where a theta function profile in a dielectric spherical shell is examined. It confirms  the observation that singular profiles show weaker divergences than smoother, less singular profiles.    

\section{Acknowledgements}

I thank M. Bordag for advice.

\section{Appendix A: Asymptotic expansions of the modified Bessel functions}

The expansion in negative powers of the parameter $l$ (eq. (24)) is obtained from the following expansions
\begin{equation}
I_{l+a}(kR) \ \sim\ \frac{1}{\sqrt{2\pi l}}\exp \{\sum_{n=-1}^3 l^{-n} S_I(n,a,t)\}\ ,
\end{equation}
\begin{equation}
K_{l+a}(kR)\ \sim\ \sqrt{\frac{\pi}{2l}}\exp \{\sum_{n=-1}^3 l^{-n} S_K(n,a,t)\}
\end{equation}
where $a$ takes the values $1$ and $0$ and the functions $S_I(n,a,t)$ and $S_K(n,a,t) $ up to the third order are given by
\begin{eqnarray}
S_I(-1,a,t) & = & t^{-1} + \frac 12 \ln\left(\frac{1 - t}{1 + t}\right)\ ,\nonumber \\
 S_I(0,a,t) & = & \frac 12\ln t -\frac{a}{2} \ln\left(\frac{1 + t}{1 - t}\right)\ ,\nonumber \\
S_I(1,a,t)& = & -\frac {t}{24} (-3 + 12a^2 + 12a t + 5 t^2)\ ,\nonumber \\
S_I(2,a,t) & = & \frac{t^2}{48} [8a^3 t + 12a^2 (-1 + 2 t^2) + 
       a(-26 t + 30 t^3) + 3 (1 - 6 t^2 + 5 t^4)]\ ,\nonumber \\
S_I(3,a,t) & = & \frac {1}{128}(((25 - 104a^2 + 16a^4) t^3)/3 + 
      16 a (-7 + 4a^2) t^4\nonumber\\
               &   & - (531/5 - 224a^2 + 16a^4)t^5 - (32a (-33 + 8a^2) t^6)/3\nonumber \\
               &   & - (-221 + 200a^2) t^7 - 240a t^8 - (1105 t^9)/9)\ ;
\end{eqnarray}
\begin{eqnarray}
S_K(-1,a,t) & = & -t^{-1} -\frac 12 \ln\left(\frac{1 - t}{1 + t}\right)\ ,\nonumber \\
 S_K(0,a,t) & = & \frac 12\ln t +\frac{a}{2} \ln\left(\frac{1 + t}{1 - t}\right)\ ,\nonumber \\
S_K(1,a,t)& = & -\frac {t}{24} (-3 + 12a^2 + 12a t + 5 t^2)\ ,\nonumber \\
S_K(2,a,t) & = & \frac{t^2}{48} [-8a^3 t + 12a^2 (-1 + 2 t^2) + 
       a(-26 t + 30 t^3) + 3 (1 - 6 t^2 + 5 t^4)]\ ,\nonumber \\
S_K(3,a,t) & = & -\frac {1}{128}((25 - 104a^2 + 16a^4) t^3)/3 + 
      16 a (-7 + 4a^2) t^4\nonumber\\
               &   & - (-531/5 + 224a^2 - 16a^4)t^5- (32a (-33 + 8a^2) t^6)/3\nonumber\\
               &   & - (221 - 200a^2) t^7 - 240a t^8 + (1105 t^9)/9\ ,
\end{eqnarray}
with $t=(1+(kR/l)^2)^{-\frac 12}$.

\section{Appendix B: Calculation of the sum and of the integrals}

The transformation of the summation over $l$  in (26) into an integral is done with the Abel-Plana formula

\begin{equation}
\sum_{l=1}^\infty F(l)\ =\ \int_0^\infty dl\ F(l)\ \ -\ \frac 12 F(0)\  +\ \int_0^\infty\frac{dl}{1-e^{2\pi\nu}}\frac{F(il)-F(-il)}{i}\ .
\end{equation}
In our case the function $F(l)$ is 
\begin{equation}
F(l)=\int_{m}^\infty dk (k^2+m^2)^{1-s}\partial_k \left(\sum_{n=1}^3\sum_jX_{n,j}\frac{t^j}{l^n}\right) \ .
\end{equation}
To the first addend of eq.(46) corresponds the contribution ${\cal E}_{as1}$ to the second addend the contribution ${\cal E}_{as2}$ ant to the third the contribution ${\cal E}_{as3}$. In ${\cal E}_{as1}$  the integrations over $k$ and $l$ are performed by means of the following formula 
\begin{equation}
\int_0^\infty dl\ \int_{m}^\infty dk\ (k^2-m^2)^{1-s}\partial_k\frac{t^j}{l^n}=-\frac{m^{2-2s}}{2}\frac{\Gamma(2-s)\Gamma\left(\frac{1+j-n}{2}\right)\Gamma(s+\frac{n-3}{2})}{(Rm)^{n-1}\Gamma(j/2)}\ .
\end{equation}
giving directly result (29). For the integration over $k$ in ${\cal E}_{as2}$ and ${\cal E}_{as3}$ the following formula has been used
\begin{equation}
\int_{m}^\infty dk\ (k^2-m^2)^{1-s}\partial_k\frac{t^j}{l^n}=-m^{2-2s}\frac{\Gamma(2-s)\Gamma\left(s+\frac j2-1\right)l^{j-n}}{\Gamma \left(\frac j2\right) (Rm)^{j}\left( 1+\left(\frac{l}{mR}\right)^2\right)^{s+\frac j2-1}}\ ,
\end{equation}
which, applied to ${\cal E}_{as2}$, leads directly to result (30). In ${\cal E}_{as3}$ the integrand is more complicated and the integrals over $l$ cannot be analytically calculated. Result (31) represents the maximum possible simplification, which we reached  after  several partial integrations. The functions $h_i(x)$ are
\begin{eqnarray}
h_1(x) & = & \int_{x}^{\infty} \frac{dl}{1-e^{2\pi l}} \sqrt{l^2 -x^2} \nonumber \\
h_2(x) & = & \int_{x}^{\infty} dl \left(\frac{1}{1-e^{2\pi l}}\frac 1l \right)' (l^2 -x^2) \nonumber \\
h_3(x) & = & \int_{x}^{\infty} dl \left(\frac{1}{1-e^{2\pi l}}\frac 1l \right)' \sqrt{l^2 -x^2} \nonumber \\
h_4(x) & = & \int_{x}^{\infty} dl \left(\left(\frac{l}{1-e^{2\pi l}}\right)'\frac 1l \right)' \sqrt{l^2 -x^2} \nonumber \\
h_5(x) & = & \int_{x}^{\infty} dl \left( \left( \left(\frac{l^3}{1-e^{2\pi l}}\right)'\frac 1l \right)' \frac 1l\right)' \sqrt{l^2 -x^2}\ .
\end{eqnarray}

\vfill \eject

\begin{figure}[ht]\unitlength1cm
\begin{picture}(6,6)
\put(-0.5,0){\epsfig{file=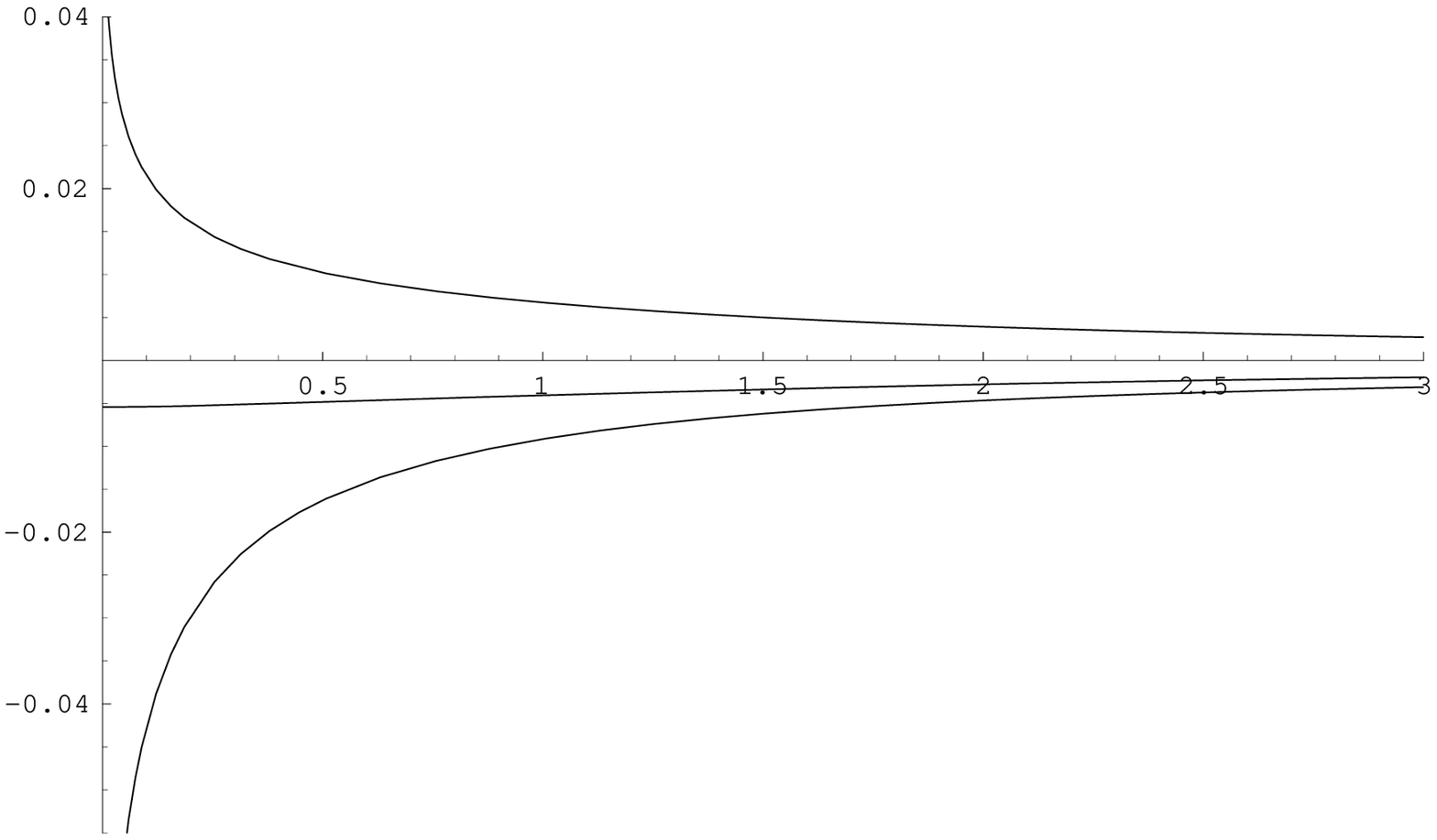,width=14cm,height=6cm}}
\put(1.4,2.7){${\cal E}_{fl} $}
\put(0.8,5.6){${\cal E}_{f0} $}
\put(0.9,0.3){${\cal E}_{as}  $}
\put(13.7,3.3){$R$}
\end{picture}
\caption{Repulsive potential. The contributions to the  renormalized vacuum energy multiplied by $R^{2}\cdot \alpha^{-2}$, for $\alpha=2.1$ .} 
\end{figure}

\begin{figure}[ht]\unitlength1cm
\begin{picture}(6,6)
\put(-0.5,0){\epsfig{file=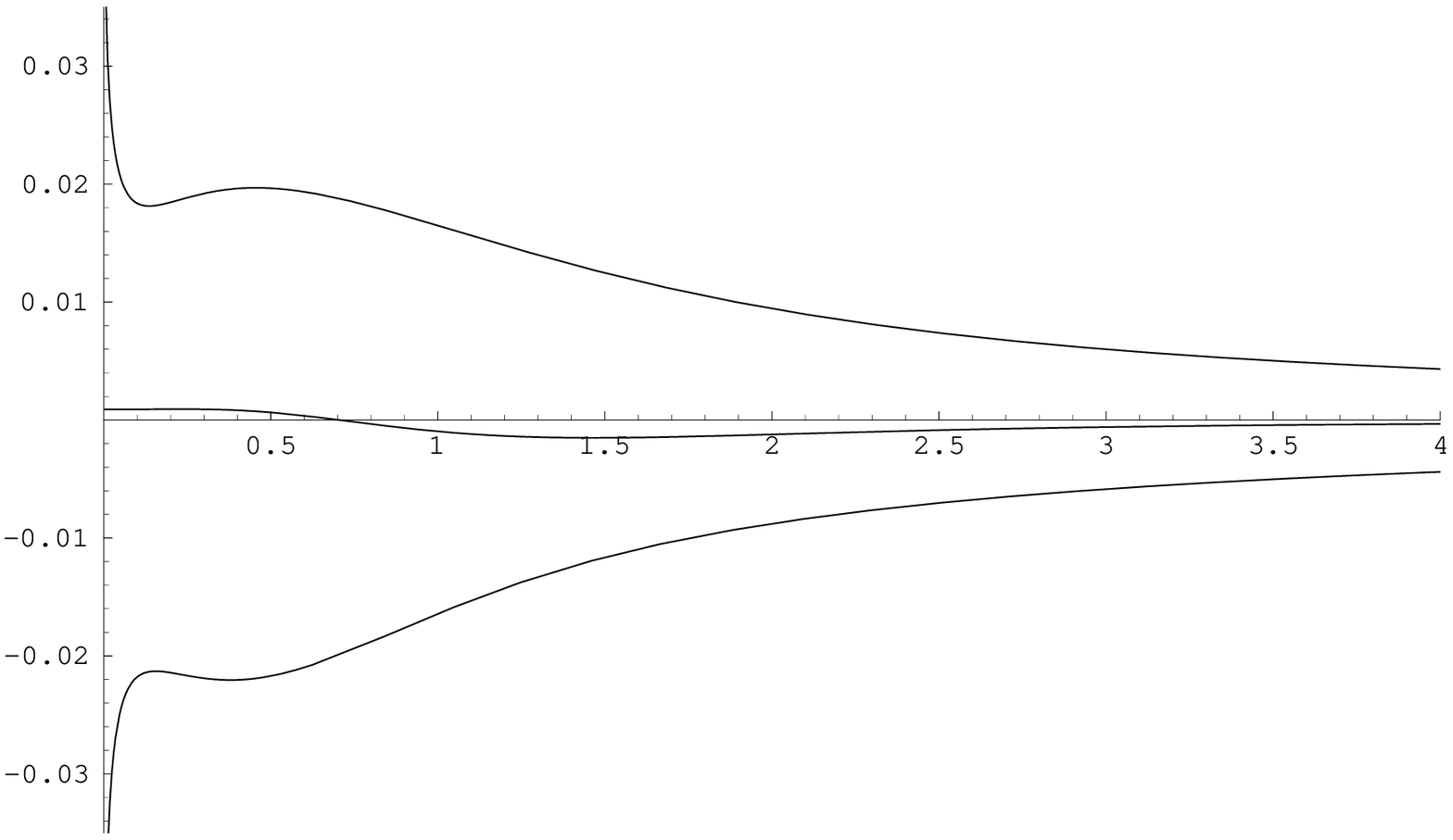,width=14cm,height=6cm}}
\put(1.6,0.7){${\cal E}_{as} $}
\put(1.6,4.9){${\cal E}_{f0} $}
\put(1.2,3.25){${\cal E}_{fl} $}
\put(13.7,2.9){$R$}
\end{picture}
\caption{Repulsive potential. The contributions to the  renormalized vacuum energy multiplied by $R^{2}\cdot \alpha^{-2}$, for $\alpha=0.3$ .} 
\end{figure}

\begin{figure}[ht]\unitlength1cm
\begin{picture}(6,6)
\put(-0.5,0){\epsfig{file=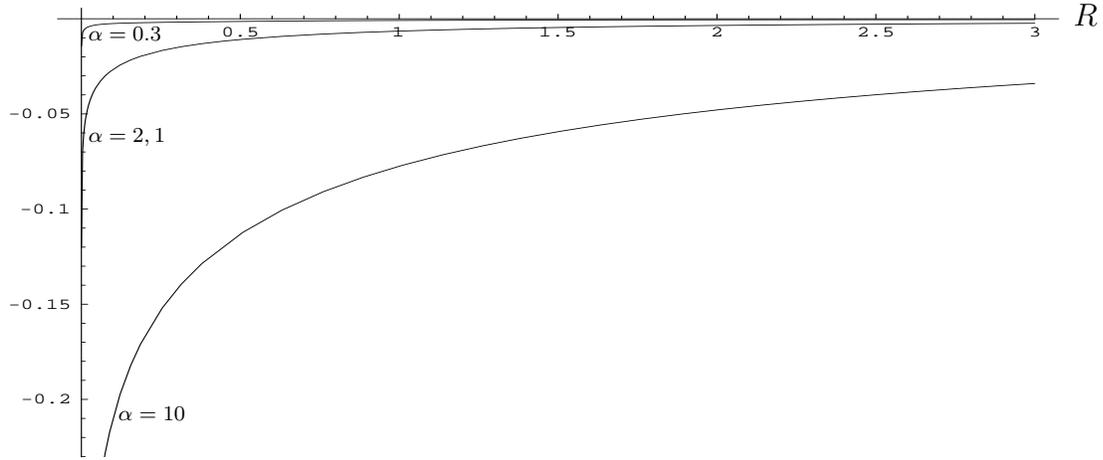,width=14cm,height=6cm}}
\put(1.0,0.5){{\scriptsize $\alpha=10$}}
\put(0.6,5.55){{\scriptsize $\alpha=0.3$}}
\put(0.6,4.2){{\scriptsize $\alpha=2,1$}}
\put(13.7,5.75){$R$}
\end{picture}
\caption{Repulsive potential. The complete renormalized vacuum energy  ${\cal E}_{ren}(R)$ multiplied by $R^{2}\cdot \alpha^{-2}$, for $\alpha=0.3$, $\alpha=2.1$ and $\alpha=10$.} 
\end{figure}

\begin{figure}[ht]\unitlength1cm
\begin{picture}(6,6)
\put(-0.5,0){\epsfig{file=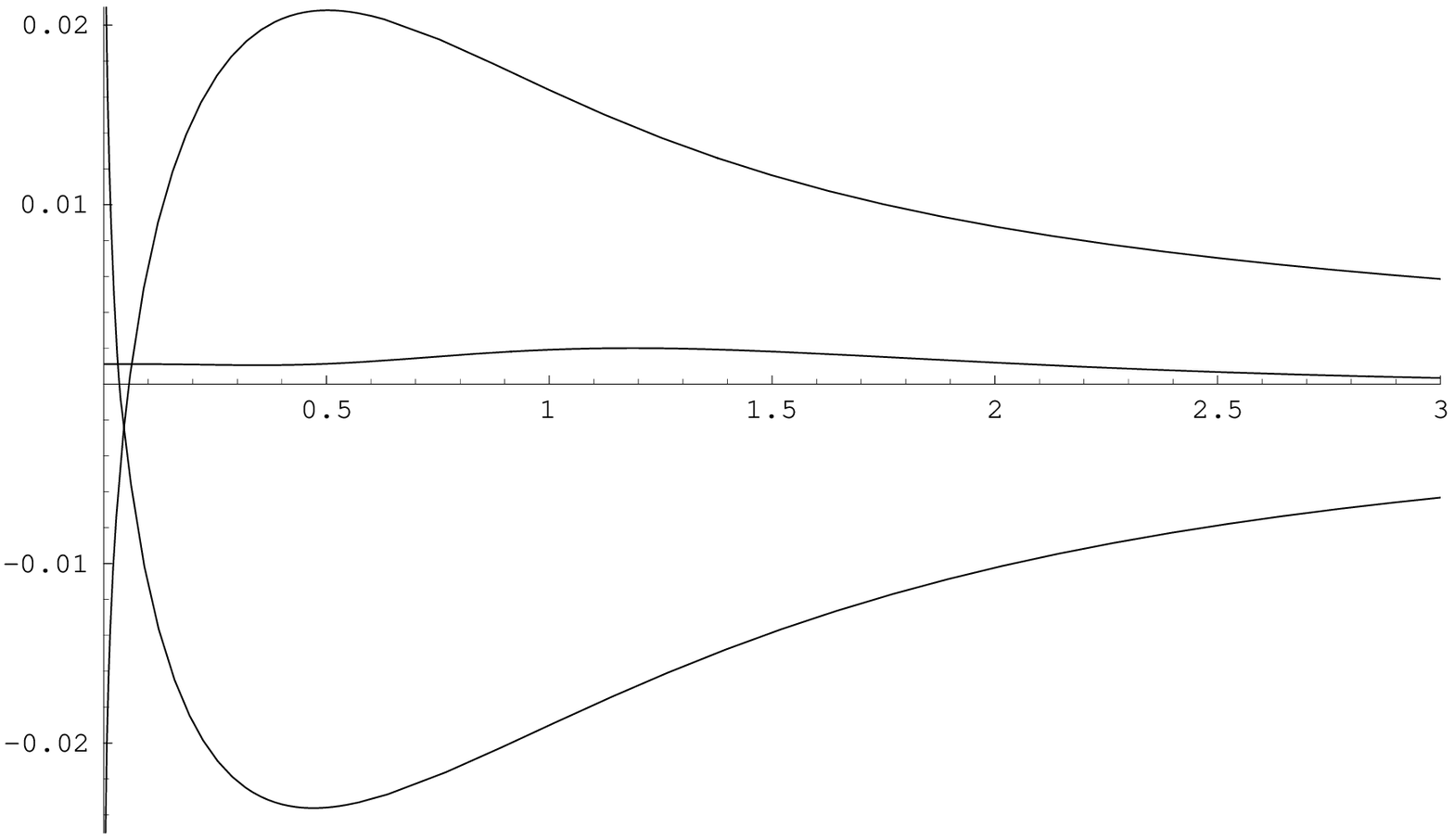,width=14cm,height=6cm}}
\put(5.6,0.7){${\cal E}_{as} $}
\put(4.4,5.7){${\cal E}_{f0} $}
\put(2.5,3.6){${\cal E}_{fl} $}
\put(13.7,3.2){$R$}
\end{picture}
\caption{Attractive potential. The contributions to the  renormalized vacuum energy multiplied by $R^{2}\cdot \alpha^{-2}$, for $\alpha=-0.3$ .} 
\end{figure}

\begin{figure}[ht]\unitlength1cm
\begin{picture}(6,6)
\put(-0.5,0){\epsfig{file=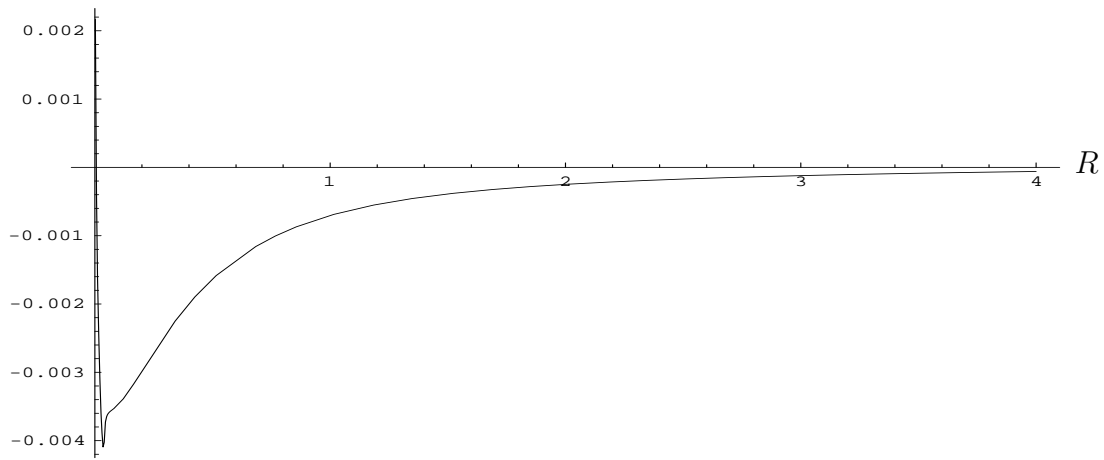,width=14cm,height=6cm}}
\put(13.7,3.8){$R$}
\end{picture}
\caption{Attractive potential. The complete renormalized vacuum energy  ${\cal E}_{ren}(R)$ multiplied by $R^{2}\cdot \alpha^{-2}$, for $\alpha=-0.3$. The peak beneath the lowest point of the curve is due to the presence of a small imaginary part in the renormalized energy causing the function to be no more analytical. The energy starts to be complex at $R \sim 0.04$.} 
\end{figure}

\end{document}